\documentclass[aps,prl,twocolumn,reprint,superscriptaddress]{revtex4-1}

\usepackage{graphicx}

\begin{document}

\title{Hybridization Induced Transparency in composites of metamaterials and atomic media}

\author{P. Weis}

\affiliation{Department of Physics and Research Center OPTIMAS, University of Kaiserslautern, Erwin-Schroedinger-Strasse, 67663 Kaiserslautern, Germany}

\author{J. L. Garcia-Pomar}

\affiliation{Department of Physics and Research Center OPTIMAS, University of Kaiserslautern, Erwin-Schroedinger-Strasse, 67663 Kaiserslautern, Germany}
\affiliation{Fraunhofer Institute for Physical Measurement Techniques IPM, Heidenhofstrasse 8, 79110  Freiburg, Germany}

\author{R. Beigang}

\affiliation{Department of Physics and Research Center OPTIMAS, University of Kaiserslautern, Erwin-Schroedinger-Strasse, 67663 Kaiserslautern, Germany}
\affiliation{Fraunhofer Institute for Physical Measurement Techniques IPM, Heidenhofstrasse 8, 79110  Freiburg, Germany}

\author{M. Rahm}
\email[]{mrahm@physik.uni-kl.de}
\homepage[]{http://www.physik.uni-kl.de/rahm}

\affiliation{Department of Physics and Research Center OPTIMAS, University of Kaiserslautern, Erwin-Schroedinger-Strasse, 67663 Kaiserslautern, Germany}
\affiliation{Fraunhofer Institute for Physical Measurement Techniques IPM, Heidenhofstrasse 8, 79110  Freiburg, Germany}


\begin{abstract}
We report hybridization induced transparency (HIT) in a composite medium consisting of a metamaterial and a dielectric material. We develop an analytic model that explains HIT by coherent coupling between the hybridized local fields of the metamaterial and the dielectric or an atomic system in general. In a proof-of-principle experiment, we evidence HIT in a split ring resonator metamaterial that is coupled to $\alpha$-lactose monohydrate. Both, the analytic model and numerical full wave simulations confirm and explain the experimental observations. HIT can be considered as a hybrid analogue to electromagnetically induced transparency (EIT) and plasmon-induced transparency (PIT).
\end{abstract}


\maketitle

\section{Introduction}

In the past decade, the research field of metamaterials encompassed a great variety of research topics ranging from fundamental to application-oriented research. Especially nonlinear metamaterials \cite{zharov2003, smith2010b}, metamaterials with gain \cite{spaser,Wegener2008, xiao}, actively tunable metamaterials \cite{chen2006,chen2007,Paul2009}, transformation optics \cite{cloak} and quantum metamaterials \cite{PIT, Liu2009,Liu2011,Artar2011,SDLiu2011,Bozh2011} lay the foundation for the development of ground-breaking photonic devices while they concurrently provide a rich nutrient medium for systematic studies of fundamental physical effects. In the latter context, metamaterials can serve as important, artificially created model systems for the investigation of novel phenomena that cannot occur in natural materials. In particular, they provide a suitable platform to study important analogies between basic physical properties that occur in the realms of different disciplines as e.\ g.\ analogue effects in classical and quantum systems \cite{PIT,Liu2009,Liu2011,Liu2009b}, electromagnetic and acoustic/mechanical systems \cite{cloak,cummer2007}, etc. In this respect, it was recently shown that metamaterials can be deliberately devised to exhibit electromagnetic properties that can be explained by the underlying concepts of electromagnetically induced transparency (EIT) in atomic systems \cite{Boller1991}. In EIT, a spectrally narrowband transmission window within a broad absorption band is established by destructive interference between the transitions of atomic quantum states \cite{Fleis}. In the light of a simplified explanation and exemplarily employing the so called $\Lambda$-configuration, EIT involves three different quantum states $|a\rangle$, $|b\rangle$ and $|c\rangle$. Hereby, the transitions between states $|a\rangle$ and $|b\rangle$ and between $|c\rangle$ and $|b\rangle$ are dipole allowed while state $|c\rangle$ is metastable with respect to a transition between $|c\rangle$ and $|a\rangle$. In this system, population can be transferred from $|a\rangle$ to $|b\rangle$ either by broadband absorption of an external probe field that drives the direct path $|a\rangle$- $|b\rangle$ or alternately by the pathway $|a\rangle$-$|b\rangle$-$|c\rangle$-$|b\rangle$. In the latter case, a dressing laser is required to coherently couple the quantum states $|c\rangle$ and $|b\rangle$. Under proper conditions, the transition paths interfere destructively and a spectrally narrowband transmission window opens up within the broadband absorption line.
\begin{figure}[b]
\includegraphics[width=0.7\columnwidth]{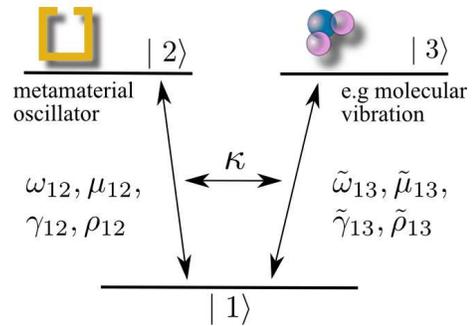}
\caption{\label{fig:levels} Level scheme of a hybrid system composed of a plasmonic metamaterial and an atomic media. The two resonators have similar resonance  frequencies $\omega_{12}\approx\tilde{\omega}_{13}$ (further properties as described in the text) couple through the local electric fields represented by the coupling strength $\kappa$.}
\end{figure}
In 2008, Zhang et al.\ theoretically described plasmon-induced transparency (PIT) in a metamaterial, an effect that strongly resembles EIT in atomic systems \cite{PIT}.  This analogue phenomenon relies on destructive interference between a bright dipole mode and a dark quadrupole mode in artificially designed plasmonic structures that mimic the role of artificial molecules. In 2009, Liu et al.\ experimentally validated PIT in a plasmonic metamaterial \cite{Liu2009}. The main advantage of dealing with PIT in comparison to EIT is the ability to deliberately tailor the total near- and farfield electromagnetic response of the plasmonic system on a subwavelength scale. Moreover, PIT no longer requires a dressing laser. Although it is possible from the principles to prepare the quantum states of an atomic system within certain limits, this task is much more involved in comparison with the design and fabrication of plasmonic molecules. On the other hand, the maximal achievable transparency and dispersion in both systems is determined by the ratio between the resonance width of the narrowband and the broadband mode. However, in PIT this ratio is inherently limited by intrinsic metallic loss of the plasmonic molecule and thus the effect of PIT is usually less pronounced than EIT in atomic systems.
In view of these limitations, it would be desirable to combine the advantages of both worlds. Therefore, the question arises if induced transparency can also be achieved in a hybrid system composed of plasmonic structures and an ensemble of atoms. The latter could be a quantum structure or a simple dielectric with extremely narrowband absorption lines. Furthermore, the plasmonic molecules could be deliberately designed for spectrally broadband radiative dipole coupling to external fields. If the mutual interaction between the plasmonic and atomic system could be controlled such that the broadband oscillations in the plasmonic molecules interacted with the narrowband oscillations of the atomic system to interfere destructively, induced narrowband transparency within the broadband absorption line would become observable. In such a system, the EIT-like effect would be rooted in the hybridized fields of the plasmonic molecules and the atomic system. Therefore, we refer to this effect as hybridization induced transparency (HIT) throughout the text.
In this article, we describe and discuss the effect of HIT on the basis of an analytic model and substantiate the validity of the analytic description by numerical full wave simulations. Even more importantly, we evidence HIT in a composite structure consisting of split ring resonators (SRRs) and $\alpha$-lactose monohydrate. In this proof-of-principle experiment, HIT is observed due to the coherent coupling of a broadband mode of the SRRs to a narrowband mode resonance of $\alpha$-lactose. Although here we restrain ourselves to the hybridization of a metamaterial and an atomic/molecular system, it is important to notice that HIT is a general effect that is based on the mode hybridization of a set of oscillators. Hence, the atomic system could be represented by a wide range of solids, fluids and gaseous materials and could open new routes for the investigation of the coupling between metamaterials and ultracold gases. Moreover, the atomic system could be substituted by quantum structures \cite{Walther2010, Wu2010, Dietze2011}, mechanical oscillators \cite{Teufel2011}, phonon excitations \cite{Neub2010, Shelton2011} or waveguides \cite{Linden2001, Zhang2009, Giess2010,Tang2011}, to name only a few examples that can be described in the regime of HIT.

\section{Analytic Model}

\begin{figure}[t]
\includegraphics[width=\columnwidth]{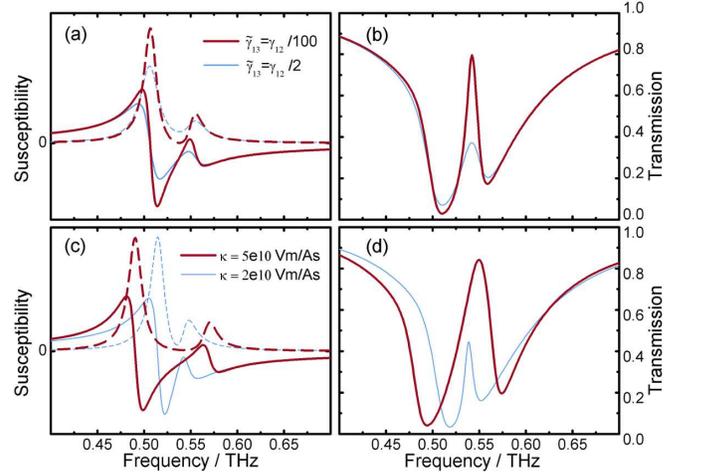}
\caption{\label{fig:suscept} Susceptibility (left column) and transmission (right column) of the hybrid system for several ratios $\gamma_{12}/\tilde\gamma_{13}$ (upper row) and coupling strength $\kappa$(lower row).}
\end{figure}

The analytic model is based on a density matrix formalism and the solution of Maxwell-Bloch equations as described in detail in \cite{Wegener2008}. In a very simplified approach, we can describe the plasmonic and the atomic system as 2-level oscillators. This is illustrated in the level diagram of Fig.\ref{fig:levels} . In this picture, the density matrix $\rho_{12}$ of the plasmonic system and $\tilde{\rho}_{13}$ of the atomic system obey the coupled differential equations
\begin{eqnarray}
\dot{\rho}_{12} & = & -i\left(\Delta_{12}-i\gamma_{12}\right)\rho_{12}+
i\frac{\mu_{12}}{\hbar} E_L \label{eq:rho12} \\
\dot{\tilde{\rho}}_{13} & = &  -i\left(\tilde{\Delta}_{13}-i\tilde{\gamma}_{13}\right)\tilde{\rho}_{13}+
i\frac{\tilde{\mu}_{13}}{\hbar}\tilde{E}_L \label{eq:rho13}
\end{eqnarray}
Hereby, $\Delta_{12}=\omega_{12}-\omega$ and $\tilde{\Delta}_{13}=\tilde{\omega}_{13}-\omega$ denote the detuning between the resonance frequencies $\omega_{12}$, $\tilde{\omega}_{13}$ and the excitation frequency $\omega$, $\mu_{12}$ and $\tilde{\mu}_{13}$ are the transition dipole moments, $\gamma_{12}$ and $\tilde{\gamma}_{13}$ are the damping constants and $E_L=E_{ext}+\kappa \tilde{P}_{13}$ and $\tilde{E}_L=E_{ext}+\kappa P_{12}$ are the local electric fields that are induced by the external electric field $E_{ext}$ and the mutually excited polarizations $P$ and $\tilde{P}$. The parameter $\kappa$ describes the coupling strength between the plasmonic and atomic oscillators. The macroscopic polarizations can be calculated by
\begin{eqnarray}
P_{12} & = & N\mu_{12}\rho_{12} \label{eq:P12}\\
\tilde{P}_{13} & = & \tilde{N}\tilde{\mu}_{13}\tilde{\rho}_{13}  \label{eq:P13}
\end{eqnarray}
$N$ and $\tilde{N}$ describe the number densities of plasmonic and atomic oscillators. From the solutions of Eqs.\ (\ref{eq:rho12})--(\ref{eq:P13}) and by homogenizing the composed system by a simplified Maxwell-Garnett approach , we obtain for the complex susceptibility $\chi$ of the composed system:
\begin{eqnarray}
\chi & = & \frac{1}{2\epsilon_0}\left( \frac{\tilde{P}_{13}}{E_{ext}}+\frac{P_{12}}{E_{ext}}\right)\label{eq:chi}
\end{eqnarray}
For simplification of the following discussion, it is justified to assume that $\omega_{12}\approx \tilde{\omega}_{13}$, which means that the resonance frequencies of the transitions $|1\rangle \rightarrow |2\rangle$ and $|1\rangle \rightarrow |3\rangle$ are approximately equal. From the steady-state solutions of (\ref{eq:rho12}) and (\ref{eq:rho13})  we obtain by applying (\ref{eq:chi}) for the imaginary part of the susceptibility $\chi''$ at frequency $\omega_{12}\approx \tilde{\omega}_{13}$ for $\Delta_{12}=\tilde{\Delta}_{13}=0$:
\begin{eqnarray}
\chi'' & = & \frac{1}{2\epsilon_0\hbar}\left(\frac{1}{A_{12}^{-1}+\frac{1}{\hbar^2}
\kappa^2\tilde{A}_{13}}+\frac{1}{\tilde{A}_{13}^{-1}+\frac{1}{\hbar^2}
\kappa^2A_{12}} \right) \label{eq:imagchi}
\end{eqnarray}
with $A_{12}=\frac{N\mu_{12}^2}{\gamma_{12}}$ and $\tilde{A}_{13}=\frac{\tilde{N}\tilde{\mu}_{13}^2}{\tilde{\gamma}_{13}}$.
By inspection of (\ref{eq:imagchi}) we observe that, for a given $A_{12}$, the imaginary part of the susceptibility $\chi''$ and thus the absorption is reduced to a minimum (I) if $\tilde{A}_{13}$ is maximized  and (II) if the coupling constant $\kappa$ is maximized. In the first case, $\tilde{A}_{13}$ can be maximized by exploiting a spectrally narrowband resonance ($\tilde{\gamma}_{13}$ small) in the atomic system that overlaps with the broadband resonance of the metamaterial. Fig.\ \ref{fig:suscept} shows the real and imaginary part of the susceptibility and the transmission of a hybrid system for two different ratios $\tilde{\gamma}_{13}/\gamma_{12}$ (Figs.\ \ref{fig:suscept}(a),(b)) and coupling factors (Figs.\ \ref{fig:suscept}(c),(d)) which confirm the discussed behavior.

\section{Experiment and Simulation  }

\begin{figure*}
\includegraphics[width=\textwidth]{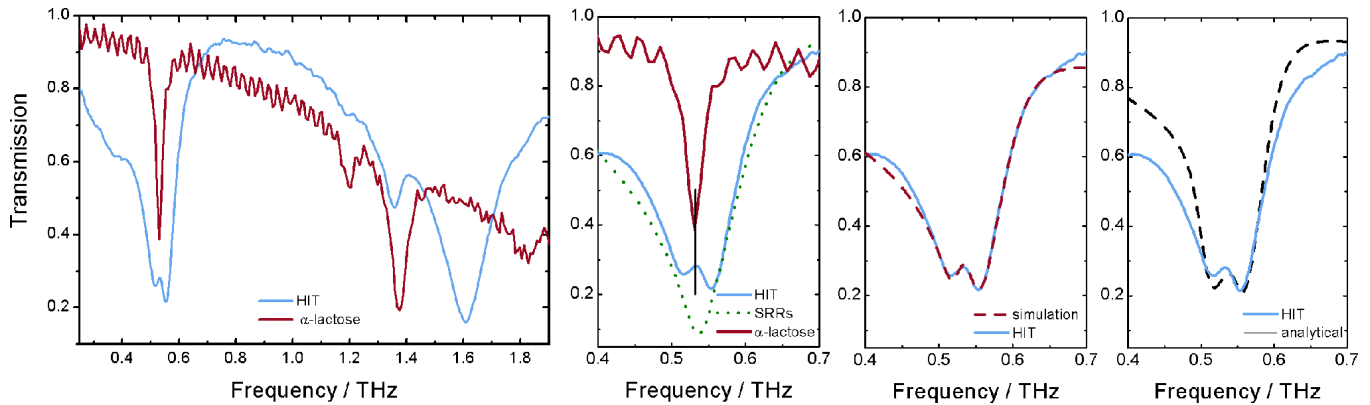}
\caption{\label{fig:spectra} Transmission spectra. (a) measurements of an $\alpha$-lactose sample (black) and the hybrid system (blue). (b) enlargement of the resonant frequency region and spectrum of the metamaterial sample without $\alpha$-lactose layer (green dotted). For a better comparison this curve has been shifted to match the resonance frequency of the lactose covered system. (c)/(d) compares the experimental to the simulative / analytical results.}
\end{figure*}
In order to verify our model, we investigated HIT in the coherent coupling between a narrowband absorption resonance of $\alpha$-lactose monohydrate and the broadband resonance of an array of SRRs in a proof-of-principle experiment. $\alpha$-lactose exhibits a spectrally narrowband, strong resonance at a frequency of 0.53\,THz \cite{ Allis2007203, Brown2007,NJPlactose}. The metamaterial structure was composed of a single layer of square-shaped gold SRRs on top of a 27\,$\mu$m thick benzocyclobutene (BCB) substrate. The SRRs had a thickness of 200\,nm, a side length of 58\,$\mu$m and the split gap was 7\,$\mu$m. The unit cells were arrayed with a periodicity of 75\,$\mu$m. For studying the coupling between the SRR metamaterial and an atomic system, we coated the SRRs with a 50\,$\mu$m thick layer of $\alpha$-lactose. To obtain a thin lactose film we diluted $\alpha$-lactose powder in a saturated aqueous solution and deposited it on top of the metamaterial. To accelerate the crystallization process, we heated the sample to a temperature of 90$^\circ$\,C, thus minimizing mutarotation of lactose from the $\alpha$ to the $\beta$ anomeric form \cite{lefort2006}.
We measured the transmission spectrum of the SRR-lactose sample by terahertz (THz) time domain spectroscopy. The polarization vector of the THz electric field was oriented along the gap of the SRRs in order to excite the magnetic SRR resonance at normal incidence by electro-magnetic cross coupling. The amplitude transmission spectrum is shown in Fig.\ \ref{fig:spectra}(a). Globally, the spectrum of $\alpha$-lactose displays three distinct transmission minima. The minima at 0.53\,THz, 1.2\, THz and 1.37\,THz are related to absorption lines of $\alpha$-lactose. The composed metamaterial/lactose system reveals four transmission minima whereof the minima at 1.2\, THz and 1.37\,THz purely originate from $\alpha$-lactose and the minimum at 1.6\,THz corresponds to the Mie resonance of the SRRs \cite{Enkirch05}. The broadband transmission minimum around 0.53\,THz stems from the magnetic SRR resonance. Taking a closer look at the minimum in Fig.\ref{fig:spectra}(b), we observe that the broadband absorption is superimposed by a spectrally narrowband transmission window within the broadband SRR resonance at a frequency of 0.53\,THz. This frequency is similar to the resonance frequency of the narrowband absorption of $\alpha$-lactose. The induced transparency traces back to the coherent, destructively interferent coupling between the broadband SRR resonance oscillations and the narrowband, absorptive transition in $\alpha$-lactose. In this respect, the SRR metamaterial and the $\alpha$-lactose interact via the hybridized local electric fields.
To substantiate this explanation, we further investigated the interaction process between the SRR metamaterial and $\alpha$-lactose by 3D full wave simulations based on a Finite Integration Technique in the frequency domain (CST Microwave Studio). In the model, the gold SRRs were described as lossy metal. The permittivity of the BCB was 2.67 with a loss tangent $\tan{\delta}=0.1$.  The geometric dimensions of the SRRs were identical to the experimental values. We described the permittivity of the narrowband absorption resonance of $\alpha$-lactose by a Lorentz-model
\begin{equation}
\epsilon=\epsilon_b+ \frac{\tilde{b}_{13}}{\tilde{\omega}_{13}^2-i\tilde{\gamma}_{13}\omega-\omega^2}
\end{equation}
Hereby, $\epsilon_b$ denotes the off-resonance background permittivity of lactose, $\tilde{\omega}_{13}$ is the resonance frequency of the transition, $\tilde{\gamma}_{13}$ is the damping factor and $\tilde{b}_{13}$ is the oscillator strength per s$^2$. We obtained a value of $\epsilon_b=3.01$ and a resonance frequency $\tilde{\omega}_{13}=2\pi\cdot0.53$\,THz from spectroscopic measurements of $\alpha$-lactose on a BCB substrate. From the spectral width of the absorption line of $\alpha$-lactose we deduced a damping factor of the resonance $\tilde{\gamma}_{13}=1.59\cdot10^{11}$s$^-1$, whereas we employed the lactose density dependent oscillator strength per s$^2$ as a free parameter to fit the simulation results to the experimental data. From this optimization we obtained a value $\tilde{b}_{13}=2.24\cdot10^{23}$s$^{-2}$.
The calculated transmission spectrum is shown by the dashed line in Fig.\ \ref{fig:spectra}(c). The transmission spectrum is in good agreement with the experimentally observed transmission and the HIT effect can be well reproduced by the numerical calculations. The simulation results indicate coherent coupling between the SRR resonance and the lactose resonance and evidence the effect of HIT.
To further corroborate the physical interpretation of HIT, we employed the analytic model to determine the transmission spectrum. For this purpose, we calculated the imaginary part of the susceptibility of Eq.\ \ref{eq:chi} by solving the coupled differential equations (\ref{eq:rho12}) and (\ref{eq:rho13}). For the model, we used the SRR and lactose resonance frequencies and the damping factors from the experimental data while we retrieved the dipole moments and the number densities from numerical calculations. We optimized the coupling factor $\kappa$ to fit the analytic transmission spectrum to the experimental data. As can be seen from Fig.\ \ref{fig:spectra}(d), the analytic model accurately describes the transmission spectrum of the composite metamaterial/lactose system. Since the model is based on coherent coupling of the hybridized local fields of a metamaterial and an atomic system, the good agreement with the experimental measurements and the numerical simulations provides a convincing proof of HIT. Although only demonstrated for a specific proof-of-principle experiment, HIT is a general effect that describes coherent coupling in hybrid structures composed of plasmonic and atomic oscillators. Since -- in principle -- the optical properties of the plasmonic system and the atomic system can be designed almost independently, HIT provides an enormous freedom for the deliberate design of optical systems with tailored dispersion.
Recently a similar experiment has been reported by Hutchison et al., who discovered a narrow transmission window in a hole array structure away from its resonance frequency when covered by a J-aggregate at the absorption frequency of the dye  \cite{Hutch2011}. They traced this observation back to interactions between the localized surface plasmons and the molecular absorption and called this effect Absorption-Induced Transparency. This experiment can be interpreted in the light of the analytical model presented in this paper, where we demonstrated that the induced transparency originates from coherent interactions between the hybridized fields of two or more oscillating systems that can be different in nature (mechanical, electromagnetic, acoustic, etc.) rather than from pure absorption of one of the oscillating systems. For this reason, we conclude that the term hybridization induced transparency adequately describes the underlying physical mechanism of a great variety of considered systems involving induced transparency phenomena.

\section{Conclusion}

In conclusion, we showed that hybridization induced transparency (HIT) can occur in a composite consisting of a metamaterial and an atomic medium. HIT is the hybrid analogue of electromagnetically induced transparency (EIT) and plasmon-induced transparency (PIT) in a mixed plasmonic/atomic structure. We explained HIT based on an analytic model that describes the coherent coupling between the oscillations of plasmonic and atomic oscillators. In a proof-of-principle experiment, we evidenced HIT in a hybrid structure composed of a split ring resonator metamaterial and $\alpha$-lactose monohydrate. The results were substantiated by numerical full wave simulations. HIT is a general effect that can be exploited for tailoring the dispersion of optical systems that consist of a mixture of plasmonic and atomic elements at will.

\section*{Acknowledgments}

We thank the Nano+Bio Center at the University of Kaiserslautern for their support in the sample fabrication and Jens Klier from the Fraunhofer Institute IPM in Freiburg for valuable discussions concerning measurement techniques. J.L. G-P. acknowledges financial support from Consolider nanolight (CSD2007-00046) and the Spanish Ministry of Education by ''Programa Nacional de Movilidad de Recursos Humanos del Plan Nacional de I-D+i 2008-2011''.

%
\end{document}